\documentclass[11pt]{article}
\usepackage{moriond,epsfig}

\def\Journal#1#2#3#4{{#1} {\bf #2}, #3 (#4)}

\def\NPB{{\em Nucl. Phys.} B}
\def\PLB{{\em Phys. Lett.}  B}
\def\PRL{\em Phys. Rev. Lett.}
\def\PRD{{\em Phys. Rev.} D}

\def\be{\begin{equation}}
\def\ee{\end{equation}}
\def\bea{\begin{eqnarray}}
\def\eea{\end{eqnarray}}

\begin{document}
\begin{flushright}
IPPP/06/33\\[-2pt]
DCPT/06/66
\end{flushright}
\vspace*{3.5cm}
\title{LEPTON FLAVOUR EFFECTS AND RESONANT LEPTOGENESIS \footnote{Talk
    presented at the XLIst Rencontres de Moriond. Based on
    work done in collaboration with A.~Pilaftsis \cite{APTU2}}}

\author{ THOMAS E.J. UNDERWOOD }

\address{Institute for Particle Physics Phenomenology, University of Durham\\
South Road, Durham DH1 3LE, United Kingdom.}

\maketitle\abstracts{
Minimal models exist in which thermal, resonant leptogenesis occurs at the
electroweak scale. The effects of individual
lepton flavours play a crucial role and allow successful leptogenesis
with large couplings between some of the charged leptons and the heavy
Majorana neutrinos. This leads to potentially observable signals for low
energy experiments (such as MEG, MECO and PRIME) and future linear colliders.
}

\section{Introduction}\label{sec:intro}
\vspace*{-0.2cm}
If the mass difference between two or more heavy Majorana neutrinos is much
smaller than their masses then the CP-asymmetry in their decays occurs
primarily through self-energy effects ($\varepsilon$-type CP-violation) rather
than vertex effects ($\varepsilon^\prime$-type CP-violation).\cite{Liu} If this
mass difference is of the order of the heavy neutrino decay widths then the
CP-asymmetry can become resonantly enhanced and may be of order 1.\cite{Pil}

This resonant enhancement can be exploited in models where the mass scale of
the heavy Majorana neutrinos is as low as the electroweak scale (i.e.
$\sim$250~GeV) allowing successful thermal leptogenesis in complete accordance
with current light neutrino data.\cite{APTU2,APTU1} Such models avoid the
severe constraints on the scale of the heavy Majorana neutrinos arising when
they have a hierarchical mass spectrum and thermal leptogenesis is required;
these constraints arise because the lightest heavy neutrino must have a mass
lower than the reheat temperature ($T_{\rm RH}$) where $T_{\rm RH}
\stackrel{<}{{}_\sim} 10^9$~GeV.  A stringent limit on the reheat temperature
is necessary in models incorporating supergravity to avoid the overproduction
of gravitinos in the early Universe, which would disrupt the nucleosynthesis
of the light elements.  In addition, the class of models under study should be
testable through experiments sensitive to lepton flavour and number violation
and may lead to the production of heavy Majorana neutrinos at a high energy
linear collider.

We will work within the framework of the Standard Model
(SM) supplemented by 3 generations of gauge-singlet right-handed neutrinos
$\nu_R$.  As in the usual see-saw scenario, these gauge singlets are permitted
to have a Majorana mass, $M_S$, violating lepton number by two units. In the
basis where the charged lepton Yukawa couplings are diagonal, the relevant
part of the Lagrangian reads 
\vspace*{-0.2cm}
\be
\label{Ynuexample}
-{\cal L}_{\rm M,Y} = \sum_{i,j=1}^3 \frac{1}{2}\,
(\bar{\nu}_{i R})^C\,(M_S)_{ij}\,\nu_{j R} \: +\:
\sum_{k=e,\mu,\tau}\, \bigg(\, \hat{h}^l_{kk}\,
\bar{L}_k\,\Phi\, l_{kR} \ +\ \sum_{l=1}^3\, h^{\nu_R}_{kl}\, \bar{L}_k\,
\tilde{\Phi}\, \nu_{lR} \bigg) \ +\ \mbox{H.c.}\,,
\vspace*{-0.2cm}
\ee
where $\tilde{\Phi}=i\tau_2\Phi^*$ is the isospin conjugate of the Higgs
doublet $\Phi$ and $\tau_2$ is the usual Pauli matrix. $L$ and $l_R$
represent lepton SU(2)$_L$ doublets and singlets respectively, $h^{\nu_R}$ is the
matrix of neutrino Yukawa couplings and $h^l$ is the matrix of charged lepton
Yukawa couplings.


\section{Lepton flavours and Resonant Leptogenesis}
\vspace{-0.2cm}
In the context of leptogenesis, the Sakharov requirement for a departure from
thermal equilibrium translates into the requirement that the heavy neutrino
decay rate is smaller than the expansion rate of the Universe $H(T)$. This can
be quantified by the parameter $K_{N_i}^l$, where $K_{N_i}^l \equiv
[\Gamma(N_i\to L_l\Phi)+\Gamma(N_i\to L_l^C \Phi^\dagger)] /H(T=m_{N_i})$,
where $\Gamma$ is the decay rate of a heavy Majorana neutrino mass eigenstate
$N_i$ with mass $m_{N_i}$.

Successful leptogenesis with hierarchical heavy neutrinos needs a relatively
small value of $K_{N_1}$, typically $K_{N_1} \sim 1$.\cite{Buchmul} A large
CP-asymmetry allows successful resonant leptogenesis with very large values of
$K_{N_i} \sim 1000$ (a very small departure from thermal equilibrium).
Conditions close to thermal equilbrium remove the dependence of the baryon
asymmetry on the initial heavy neutrino abundance or any initial lepton or
baryon asymmetries.

In general, the matrix of neutrino Yukawa couplings $h^{\nu}$ will have
different entries coupling a given heavy neutrino to each lepton flavour. This
should be taken into account by considering the dynamical generation of each
lepton flavour asymmetry. In a minimal model, an asymmetry in just one flavour
is needed to produce the baryon asymmetry of the Universe and this asymmetry
could be produced via the out of equilibrium decay of just one heavy Majorana
neutrino.\cite{PilTau} This is possible because $B+L$ violating sphaleron
interactions individually preserve $\frac{1}{3} B - L_{e,\mu,\tau}$, and will
therefore convert an individual flavour asymmetry into a baryon
asymmetry.\footnote{Several other works have studied flavour issues and
  leptogenesis.\cite{flav} In the low scale ($T<10^6$~GeV) scenarios discussed
  here, all the charged lepton Yukawa couplings are in equilibrium, favouring
  the basis in which they are diagonal. The effects of quantum oscillations
  amongst the flavour asymmetries are therefore suppressed.}

These requirements mean that $K_{N_i}$ should be relatively small for only 1
heavy neutrino $N_i$, and the $K_{N_j}^l$ for only one lepton flavour $l$
needs to be small to avoid washing out the produced asymmetry. Thus the
requirement of small Yukawa couplings, coming from demanding successful
thermal leptogenesis, can be substantially relaxed.

Once the model is specified, to accurately determine the baryon asymmetry one
must solve the relevant Boltzmann equations. It is possible to determine and
solve coupled Boltzmann equations for each lepton flavour asymmetry and also
the baryon asymmetry.\cite{APTU2} Solving a separate Boltzmann equation for
the baryon asymmetry allows the inclusion of effects due to the sphaleron
mediated processes dropping out of equilibrium at temperatures less than the
critical temperature for electroweak symmetry breaking.

If $K_{N_i} \stackrel{>}{_\sim} 1$, an order of magnitude estimate of the
baryon asymmetry may be obtained using
\vspace*{-0.3cm}
\be
\eta_B \sim -10^{-2}
\sum_{l=e,\mu,\tau}\,\sum_{i=1}^3
e^{-(m_{N_i}-m_{N_1})/m_{N_1}}\,\delta_{N_i}^l\,\frac{K_{N_i}^l}{K_l
  K_{N_i}}\,,\vspace*{-0.1cm}
\ee
where $\delta^l_{N_i}$ is the CP-asymmetry in the decay of $N_i$ into leptons of
flavour $l$, $K_{N_i} = \sum_l K^l_{N_i}$ and $K_l = \sum_i
e^{-(m_{N_i}-m_{N_1})/m_{N_1}} K^l_{N_i}$. It can be shown that the inclusion of the dynamical effects of lepton flavours
can alter the predicted baryon asymmetry in some resonant leptogenesis
scenarios by factors of order $10^{6}$. Even in scenarios with a mildly
hierarchical spectrum of heavy neutrinos the effects of individual lepton
flavours can alter predictions by a factor of 10.

\vspace{-0.3cm}
\section{A Model for Resonant {\boldmath $\tau$}-Leptogenesis}
\vspace{-0.2cm}
To fully exploit the phenomenological possibilities offered by low-scale,
resonant leptogenesis it is necessary to abandon the usual see-saw paradigm
that small light neutrino masses arise because of the large Majorana masses of
gauge singlet right-handed neutrinos.  Instead, we will consider models where
small light neutrino masses arise due to approximate flavour symmetries.
Approximate flavour symmetries can also motivate a nearly degenerate spectrum
of heavy Majorana neutrino masses.

As a definite example,\cite{PilTau} we will consider a model with a heavy
neutrino sector that is SO(3) symmetric in the absence of Yukawa
couplings.\cite{SW} This will provide 3 nearly degenerate heavy Majorana
neutrinos.  Specifically, the $\nu_{iR}$ transform as triplets under this
symmetry, whilst all other fields are singlets. Therefore, $M_S$ can be
written $M_S = m_N {\bf 1}_3 + \Delta M_S$, where $\Delta M_S$ parameterizes
deviations from the imposed SO(3) symmetry.

The presence of the charged lepton Yukawa couplings breaks the leptonic
flavour symmetry of the model down to the usual three U$(1)_{l^\prime}$
flavour symmetries of the SM and the SO(3) symmetry of the gauge singlet
neutrinos. Finally, when imposing the neutrino Yukawa couplings we choose to
leave a particular SO(2)$\simeq$U(1)$_l$ subgroup of the total flavour
symmetry unbroken (to leading order). Specifically, all lepton doublets are
coupled to the particular heavy neutrino combination:
$\frac{1}{\sqrt{2}}(\nu_{2R}+i\nu_{3R})$. The U(1)$_l$ charges of the fields
are therefore: $Q (L_{i}) = Q(l_{iR}) = 1$, $Q(\frac{1}{\sqrt{2}}(\nu_{2R}
+ i \nu_{3R}))\ =\ -Q(\frac{1}{\sqrt{2}}(\nu_{2R} - i \nu_{3R}))\ =\ 1$,
$Q(\nu_{1R})\ =\ 0$. This symmetry leads to the following generic structure for
the neutrino Yukawa couplings: \be
\label{hnur}
h^{\nu_R} \ =\ \left(\! \begin{array}{ccc}
    0  & a\, e^{-i\pi/4}  & a\, e^{i\pi/4} \\
    0  & b\, e^{-i\pi/4}  & b\, e^{i\pi/4} \\
    0 & c\, e^{-i\pi/4} & c\, e^{i\pi/4} 
\end{array} \!\right)\ +\ \left(\! \begin{array}{ccc}
    \varepsilon_e  & 0 & 0 \\
    \varepsilon_\mu  & 0 & 0 \\
    \varepsilon_\tau & 0 & 0 \end{array} \!\right)\,.
\ee
The U(1)$_l$ symmetry is broken by both $\Delta M_S$ and the second term in
equation (\ref{hnur}).\footnote{We will not consider the origin of the
  symmetry breaking terms here, but they could arise for example in a
  Froggatt-Nielsen scenario. Fine tuning is a concern, but in the examples
  considered here we have imposed a `naturalness criterion' requiring that the
  cancellation between tree and radiative effects is less than 1 part in
  $\sim20$.}

The parameters $a,b$ and $c$ in equation (\ref{hnur}) are arbitrary complex
parameters which should be smaller than $10^{-2}$ to provide a light neutrino
sector compatible with current data. In addition, they allow the choice of a
specific flavour direction in which the lepton asymmetry can be preserved. For
example, the requirement that an excess in $L_\tau$ be protected from wash-out
effects leads to the constraint $|c|\stackrel{<}{_\sim} 10^{-5}$. The small
U(1)$_l$ breaking parameters $\varepsilon_{e,\mu,\tau}$ should be of order
$10^{-6}$ allowing small light neutrino masses and one heavy neutrino to decay
sufficiently far out of thermal equilibrium. Note that the neutrino Yukawa
couplings in this model span a range comparable to that of the charged lepton
Yukawa couplings.

\begin{figure}
\begin{center}
\includegraphics[scale=0.3]{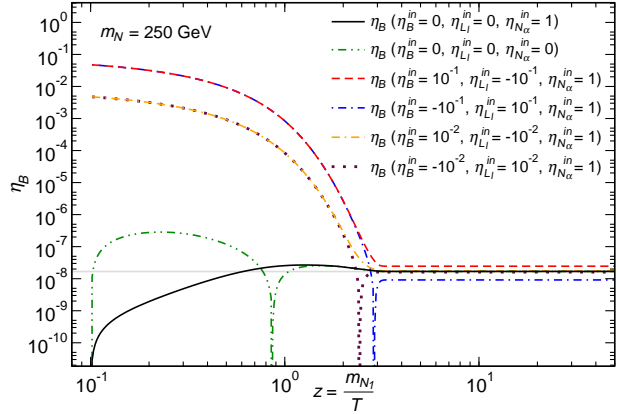}
\end{center}
\vspace{-0.5cm}
\caption{The predicted evolution of the ratio of baryon to photon number
  densities $\eta_B$, for a model with $m_N=250$~GeV. The horizontal line
  shows the value needed to agree with current observational data. The
  Boltzmann equations are solved with various initial baryon ($\eta_{B}^{\rm
    in}$) and lepton asymmetries ($\eta_{L_l}^{\rm in}$) and various initial
  heavy neutrino abundances ($\eta_{N_\alpha}^{\rm in}$).
\vspace{-0.5cm}}
\label{fig:1}
\end{figure}

Figure \ref{fig:1} shows the results of solving the Boltzmann equations with
various initial conditions in a scenario with $m_N = 250$~GeV, in complete
accordance with light neutrino data and an inverted hierarchical light
neutrino spectrum.\footnote{The Boltzmann equations were solved using the
  Fortran code {\sc LeptoGen}, available from {\tt
    http://www.ippp.dur.ac.uk/$\sim$teju/leptogen}.} The final baryon
asymmetry is almost completely independent of the initial baryon and lepton
asymmetries and the initial abundance of heavy neutrinos. In this scenario,
$|a|\sim|b|\sim10^{-2}$, $|c|\sim10^{-6}$,
$|\varepsilon_{e,\mu,\tau}|\sim10^{-6}$ and $(\Delta M_S)_{ij}/m_N$ ranged
between $10^{-5}$ and $10^{-9}$.

This choice of parameters leads to large $K_{N_{1,2}}^{e,\mu} \sim 10^{10}$,
whilst $K_{N_3}^l \sim 10$ is relatively small, allowing $N_3$ to decay
sufficiently out of thermal equilibrium to produce a lepton asymmetry. A
$\tau$-lepton asymmetry is protected from washout by $K_{N_i}^{\tau}\sim 5$.
With these $K$-factors a CP-asymmetry $\delta_{N_3}^\tau \sim -10^{-6}$ is
sufficient to generate the observed baryon asymmetry.
\vspace{-0.4cm}
\section{Phenomenology}\vspace{-0.2cm}
The observation of lepton number violation in neutrinoless double beta decay
($0\nu\beta\beta$-decay) would be an important piece of evidence pointing to
leptogenesis as the origin of the baryon asymmetry of the Universe. In the
example scenario studied, the effective Majorana mass is $|\langle
m\rangle|\simeq 0.013$~eV. This figure is close to the sensitivity of
proposals for future $0\nu\beta\beta$-decay experiments, which should at least
significantly constrain the parameter space for these models.

Heavy Majorana neutrinos with relatively large Yukawa couplings may induce
lepton flavour violating (LFV) couplings to the photon and $Z$-boson. In a
scenario with large Yukawa couplings to the electron and muon the LFV decays
$\mu\to e\gamma$ and $\mu \to eee$ can be induced. The same LFV couplings can
give rise to the coherent conversion of $\mu \to e$ in nuclei
e.g. $\mu^{-}\,{}^{48}_{22}{\rm Ti} \to e^{-}\,{}^{48}_{22}{\rm Ti}$.

In the previous example, the branching fraction for the decay $\mu \to
e\gamma$ is given by
\vspace*{-0.1cm}
\be
B(\mu \to e\gamma) = 9\times 10^{-4}\,|a|^2\,|b|^2\,v^4/m_N^4\,,
\vspace*{-0.1cm}
\ee
where $v\simeq 246$~GeV is the vacuum expectation value of the Higgs field.
For $m_N=250$~GeV we find $B(\mu \to e\gamma)\simeq 10^{-12}$, which is well
within reach of the experiment proposed by the MEG collaboration.\cite{MEG}
The branching fractions for $\mu \to eee$ and $\mu \to e$ conversion in
${}^{48}_{22}{\rm Ti}$ are related to this result, and for $m_N=250$~GeV one
obtains
\bea
B(\mu\to eee) \!&\simeq &\! 1.4 \times 10^{-2} \times B(\mu\to e\gamma)\,\,\simeq\,\,
1.4 \times 10^{-14}\,,\nonumber\\
B_{{}^{48}_{22}{\rm Ti}}(\mu \to e) \!& \simeq &\! 0.46 \times B(\mu\to
e\gamma)\,\,\simeq\,\, 4.5 \times 10^{-13}.
\eea
The result for $B_{{\rm Ti}}(\mu \to e)$ falls well within reach
of the experiments proposed by the MECO and PRIME collaborations.\cite{MECO}

Finally, in the previous example, large couplings between $N_{1,2}$ and
$e,\mu$ mean that it should be possible to produce $N_{1,2}$ at a future
$e^+e^-$ or $\mu^+\mu^-$ linear collider. Taking backgrounds into account, the
study \cite{collider} finds that an $e^+e^-$ linear collider
$\sqrt{s}=0.5$~TeV with would be sensitive to $|a| v/m_N \sim 0.7
\times 10^{-2}$.

\vspace{-0.35cm}
\section{Acknowledgements}\vspace{-0.2cm}
I would like to thank Apostolos Pilaftsis for collaboration on this project.
I would also like to thank the conference organisers and acknowledge the
financial support of a Marie Curie grant.


\vspace{-0.35cm}
\section*{References}
\begin{thebibliography}{99}\vspace{-0.2cm}\small
\bibitem{APTU2}A.~Pilaftsis, T.~E.~J.~Underwood,
  \Journal{\PRD}{72}{113001}{2005}.

\bibitem{Liu}J.~Liu and G.~Segre, \Journal{\PRD}{48}{4609}{1993}; M.~Flanz,
  E.~A.~Paschos and U.~Sarkar, \Journal{\PLB}{345}{248}{1995}; L.~Covi,
  E.~Roulet and F.~Vissani, \Journal{\PLB}{384}{169}{1996}.

\bibitem{Pil}A.~Pilaftsis, \Journal{\PRD}{56}{5431}{1997}.

\bibitem{APTU1}A.~Pilaftsis, T.~E.~J.~Underwood,
  \Journal{\NPB}{692}{303}{2004}.

\bibitem{Buchmul}W.~Buchmuller, P.~Di Bari, M.~Plumacher,
  \Journal{\NPB}{665}{445}{2003}

\bibitem{PilTau}A.~Pilaftsis, \Journal{\PRL}{95}{081602}{2005}.
  
\bibitem{flav}R.~Barbieri, P.~Creminelli, A.~Strumia, N.~Tetradis,
  \Journal{\NPB}{575}{61}{2000}; O.~Vives, \Journal{\PRD}{73}{073006}{2006};
  A.~Abada, S.~Davidson, F.~X.~Josse-Michaux, M.~Losada, A.~Riotto,
  \Journal{\em JCAP}{0604}{004}{2006}; E.~Nardi, Y.~Nir, E.~Roulet, J.~Racker,
  \Journal{\em JHEP}{0601}{164}{2006}.

\bibitem{SW}S.~M.~West, \Journal{\PRD}{71}{013004}{2005}.

\bibitem{MEG}See proposal at \texttt{http://meg.web.psi.ch}.
  
\bibitem{MECO}M.~Hebert (MECO Collaboration), \Journal{\em Nucl. Phys.
    A.}{721}{461}{2003}; K.~Yoshimura, \Journal{\em Nucl. Instrum. Meth.
    A}{503}{254}{2003}.
  
\bibitem{collider}F.~del~Aguila, J.A.~Aguilar-Saavedra,
  A.~Martinez~de~la~Ossa, D.~Meloni, \Journal{\PLB}{613}{170}{2005};
  F.~del~Aguila, J.A.~Aguilar-Saavedra, \Journal{\em JHEP}{0505}{2005}{026}.

\end{thebibliography}
\end{document}